\documentclass[10pt,journal]{IEEEtran}

\usepackage{bm}
\usepackage{tikz}
\usepackage{graphicx}
\usepackage{animate}
\usepackage{amsmath}
\usepackage{amssymb,amsthm}
\usepackage[shortlabels]{enumitem}
\usepackage{pgfplots}
\usetikzlibrary{spy, calc}
\pgfplotsset{compat=1.5.1}

\renewcommand{\vec}[1]{\ensuremath{{\bm{#1}}}} 
\newcommand{\rvar}[1]{\ensuremath{{{#1}}}} 
\newcommand{\rvec}[1]{\ensuremath{\bm{{{{#1}}}}}} 
\newcommand{\ovec}[1]{\ensuremath{\bm{{{#1}}}}} 
\newcommand{\evec}[1]{\ensuremath{{\bar{\bm{{#1}}}}}} 
\newcommand{\mat}[1]{{\ensuremath{{\mathbf{#1}}}}} 
\newcommand{\rmat}[1]{{\ensuremath{\mathbf{{{#1}}}}}} 
\newcommand{\covm}[1]{\ensuremath{\mathbf{C}^{\rvec{#1}}}} 
\newcommand{\tp}{\text{T}} 
\newcommand{\gauss}[2]{\ensuremath{\mathcal{N}\left({#1},{#2}\right)}}
\newcommand*{\Trace}[1]{\Tr\hspace{-0pt}\left(#1\right)}

\DeclareMathAlphabet{\mathcal}{OMS}{cmsy}{m}{n}
\DeclareMathOperator{\Tr}{Tr}

\begin{document}

\title{A Batch Update Using Multiplicative Noise Modelling for Extended Object Tracking}

\author{Christian Gramsch, Shishan Yang and Hosam Alqaderi
		\IEEEcompsocitemizethanks{\IEEEcompsocthanksitem C.~Gramsch and S.~Yang work at MicroVision GmbH,
           Neuer Höltigbaum~6, 22143 Hamburg, Germany (e-mail: gramsch@mailbox.org; yangshishan@gmail.com)).
           H.~Alqaderi is affiliated with the University of Bonn, Germany (e-mail: hosam.alqaderi@uni-bonn.de)\protect\\}
}

\maketitle

\begin{abstract}
  While the tracking of multiple extended targets demands for sophisticated algorithms to handle the high complexity inherent to
  the task, it also requires low runtime for online execution in real-world scenarios. In this work, we derive a batch update for
  the recently introduced elliptical-target tracker called MEM-EKF*. The MEM-EKF* is based on the same likelihood as the
  well-established random matrix approach but is derived from the multiplicative error model (MEM) and uses an extended Kalman
  filter (EKF) to update the target state sequentially, i.e., measurement-by-measurement. Our batch variant updates the target
  state in a single step based on straightforward sums over all measurements and the MEM-specific pseudo-measurements. This
  drastically reduces the scaling constant for typical implementations and indeed we find a speedup of roughly 100x in our
  numerical experiments. At the same time, the estimation error which we measure using the Gaussian Wasserstein distance
  stays significantly below that of the random matrix approach in coordinated turn scenarios while being comparable
  otherwise.
\end{abstract}


\section{Introduction}
Autonomous and assisted driving offer safer and more efficient transportation options for our society. 
With environment perception serving as a key technology enabling the realization of autonomous and assisted driving system, 
object tracking emerges as one of the most fundamental components of environment perception. 
Object tracking aims at estimating and predicting the objects states in an area of interest. 
Traditional object tracking assumes the tracked object as a point and has at most one detection. 
However, this assumption is challenged by the near-field and high-resolution sensors, which are widely equipped in modern vehicles.  
For the modern ranged-based sensors, such as RAdio detection and ranging (Radar) and LIght Detection And Ranging (Lidar), one
traffic participant emits multiple measurements and its extent cannot be reduced to a point for safety consideration. 
Extended object tracking estimates and predicts the kinematic and shape of an object given a set of detections.
A detailed overview of extended object tracking is given in \cite{granstromExtendedObject:2017}. 
  
The origins of measurements from an extended object can be modelled as reflections from specific positions on the object's surface \cite{Vermaak2005} or   an independent random draw from a spatial distribution \cite{Gilholm2005_particleFilter,Gilholm2005}, which avoids the explicit association between measurements and measurement-sources. 
The shape of an extended object can be interpreted with different levels of complexity depending on the computation and application requirements. 
For object tracking scenarios in autonomous and assisted driving, the extent of objects is often approximated by simple geometry
shapes, such as rectangles and ellipses \cite{kochBayesianApproach:2008,Granstroem2011,Lan2012b} -- 
the latter being among the most popular due to their statistical convenience in representing Gaussian distributions. 
Their tracking was pioneered by Random-matrix-based methods
\cite{kochBayesianApproach:2008,feldmannTrackingExtended:2011} which represent elliptical objects using a 2D semi-definite matrix.
While elegant, this representation causes an increased error in coordinated turn scenarios due to the implicit coupling of
the uncertainty of orientation and size.
To address this issue, a non-linear measurement model which relates measurements, position, orientation and axes-lengths was formulated in \cite{yangTrackingOrientation:2019}.
There, the key in deriving the measurement equations is the modelling via multiplicative noise -- also known as the Multiplicative
Error Model (MEM) \cite{baumModelingTarget:2012}.  
Indeed it was found that the MEM formulation improves the performance in coordinate turn scenarios significantly as the dynamics of
parameters can be constraint individually. 
However, one disadvantage of the MEM formulation \cite{yangTrackingOrientation:2019} is its sequential Kalman update, which causes a slower runtime and is sensitive to the measurement order. 

In this work, we propose a new filter for extended object tracking, which is derived from the multiplicative error model (MEM) in
\cite{yangTrackingOrientation:2019}, but uses a batch update based on the extended information filter (EIF). 
The proposed tracker is named as MEM-EIF with MEM referring to the measurement model and EIF denoting the extended information filter. 
Our new tracker MEM-EIF: i) decouples the dynamics of object orientation and axes lengths compared with Random Matrix based methods; ii) performs a batch update and thus significantly improves runtime-performance while being robust against changes in the order of processed measurements compared with MEM-EKF*. 
 
Interestingly, the information filter has been combined with the MEM-EKF* before in \cite{liDistributedExtended:2022} but in the
different context of information fusion. In \cite{liDistributedExtended:2022}, a distributed tracking system was developed which retains the consensus regarding shape and kinematics among different sensor
nodes on a network.

The rest of our paper is structured as follows. The notation is introduced in the next subsection. 
In section \ref{sec:information_filter}, we briefly explain the information filter. 
We derive our batch update using extended information filter  in Section \ref{sec:batch_update}.   
The numeric evaluation and its result is presented in Section \ref{sec:numerics}. 
The paper is concluded in Section \ref{sec:conclusion}.

\subsection{Notation}
We denote vector-valued quantities in italic, bold (e.g., $\ovec{y}=(y_1, y_2)^\tp$ for a 2d-measurement). Upright bold
quantities, such as the measurement matrix $\mat{H}$, denote matrices. Scalars are denoted in non-bold italic, e.g., $L$ for the
number of measurements. 
We indicate the expectation value of random quantities by a bar, e.g., $\evec{x} =
\mathbb{E}[\rvec{x}]$ for a vector-valued random variable $\rvec{x}$ where $\mathbb{E}[\cdot]$ denotes the operator taking the
expectation value on a suitable probability space. 
The covariance matrix of a random vector $\rvec{x}$ is denoted as $  \covm{x}$
and the matrix of the cross-covariance between two random variables $\rvec{x}$ and $\rvec{q}$ as $ \covm{xq}$.
Given $\rvec{y}_1, \dots, \rvec{y}_i$, we indicate the conditional expectation value, the conditional covariance matrix, and conditional cross-covariance  using a subscript $i$, i.e., $\evec{x}_i$, $ \covm{x}_i$, $ \covm{xq}_i$, respectively. 
Additionally, $\rvec{x}\sim\mathcal{N} \left( \bar{\rvec{x}},\covm{x} \right)$ denotes that $\rvec{x}$ is Gaussian distributed with mean $\bar{\rvec{x}}$ and covariance $\covm{\rvec{x}}$.

\section{The Information Filter}\label{sec:information_filter}
To introduce the information filter, we consider $L$
measurements \makebox{$\rvec{y}_1$, ..., $\rvec{y}_L$} subject to a linear stochastic model: 
\begin{equation}
  \label{eq:linearmodel}
  \rvec{y}_i = \mat{H} \rvec{x} + \rvec{v}_i,\text{ for } i=1,\cdots,L,
\end{equation}
where $\mat{H}$ is the measurement matrix. The measurement noise $\rvec{v}_i\sim\mathcal{N} \left( 0,\covm{v} \right)$ and the state
$\rvec{x}\sim\mathcal{N} \left(\evec{x}_0,\covm{x}_0 \right)$ are Gaussian distributed. 
For the latter, we denote $\evec{x}_0$ and $\covm{x}_0$ as the predicted mean and covariance.
Given  $\rvec{y}_1, ..., \rvec{y}_{i}$, for $i\leq L$, $\evec{x}_i$ and covariance $\covm{x}_i $ can be calculated using a sequential Kalman update
\begin{align}
  \evec{x}_i &= \evec{x}_{i-1} + \covm{xy}_i \left[ \covm{y}_i \right]^{-1} \left( \ovec{y}_i-\evec{y}_i \right),
  \label{eq:kalman_linear1}\\
  \covm{x}_i &= \covm{x}_{i-1} - \covm{xy}_i \left[ \covm{y}_i \right]^{-1} \left( \covm{xy}_i \right)^{\tp},
  \label{eq:kalman_linear2}
\end{align}
with the innovation covariance $\covm{y}_i= \covm{v} + \mat{H} \covm{x}_{i-1} \mat{H}^{\tp}$ and the cross-covariance $\covm{xy}_i= \covm{x}_{i-1}\mat{H}^{\tp}$.

As is well-known (see  \cite{thrunProbabilisticRobotics:2005}), an equivalent information filter can update with all measurements in a
single step. A brief derivation is given in Appendix~\ref{ap:info_derivation}.
We define $\evec{\xi}_i \equiv [\covm{x}_{i}]^{-1} \evec{x}_{i}$ such that
\begin{align}
  \evec{\xi}_{i} &= \evec{\xi}_{i-1} + \mat{H}^{\tp} [\covm{v}]^{-1} \ovec{y}_i,\\
  [\covm{x}_{i}]^{-1} &= [\covm{x}_{0}]^{-1} + \mat{H}^{\tp} [\covm{v}]^{-1} \mat{H}.
\end{align}
Iterating the update rule gives us the following batch update:
\begin{align}
  \label{eq:info_intro1}
  \evec{\xi}_{L} &= \evec{\xi}_{0} + \mat{H}^{\tp} [\covm{v}]^{-1} \sum_{i=1}^{L} \ovec{y}_i,\\
  [\covm{x}_{L}]^{-1} &= [\covm{x}_{0}]^{-1} + L \mat{H}^{\tp} [\covm{v}]^{-1} \mat{H},
  \label{eq:info_intro2}
\end{align}
where $\evec{x}_L = \covm{x}_L \evec{\xi}_L$ recovers the original state. 
 
The main characteristic of this batch form is the simple dependence on the outcomes $\ovec{y}_i$ which enter via their sum only. 
This summation can typically be implemented in a very efficient manner, e.g., by employing vectorization or even parallelization. 
In addition, only three matrix inversions are required to obtain the final outcome compared to $L$ such inversions in the standard Kalman form.

\section{Application to the MEM-EKF}\label{sec:batch_update}
Elliptical extend-object tracking focuses on updating the state of a target that has produced $L$ measurements. 
In a recent work \cite{yangTrackingOrientation:2019}, this setting has been modeled using the multiplicative error model
\cite{baumModelingTarget:2012} to which we give a short introduction in subsection \ref{sec:mem}. 
The closed form solution derived in \cite{yangTrackingOrientation:2019} is based on a problem-tailored sequential extended Kalman filter
(EKF). The EKF update formul\ae{} consist of two parts: one update for kinematics and another for the shape. Similarly, our batch
update consists of a kinematic and a shape update step which are derived in the subsections \ref{sec:kinematics} and
\ref{sec:shape}. In contrast to \cite{yangTrackingOrientation:2019}, however, all the measurements are processed within a single
step. 

\subsection{Multiplicative error model (MEM)}
\label{sec:mem}
We consider an elliptical target moving in the 2d-plane with unknown kinematic state $\rvec{r} = (\rvar{r}_x, \rvar{r}_y,
\rvar{r}_{\dot{x}}, \rvar{r}_{\dot{y}}, \rvar{r}_{\ddot{x}}, \rvar{r}_{\ddot{y}})^\tp$, where $(\rvar{r}_x, \rvar{r}_y)^{\tp}$
describes the target's 2d position, $(\rvar{r}_{\dot{x}}, \rvar{r}_{\dot{y}})^{\tp}$ its velocity and $(\rvar{r}_{\ddot{x}},
\rvar{r}_{\ddot{y}})^{\tp}$ its acceleration. 
The shape of the target is assumed to be that of an ellipse with unknown parameters
denoted as $\rvec{p} = (\rvar{\alpha}, \rvar{l}_1, \rvar{l}_2)$, where $\rvar{\alpha}$ is the orientation of the ellipse while
$\rvar{l}_1$ and $\rvar{l}_2$ denote the lengths of its major and minor semi-axes, respectively. 
The elliptical target is assumed to give rise to $L$ measurements $\ovec{y}_1, ..., \ovec{y}_L$ via the measurement model
\begin{align}
  \label{eq:measModel}
  \rvec{y}_i = \mat{H} \rvec{r} + \rmat{S} \rvec{h}_i + \rvec{v}_i,
\end{align}
where 
\begin{align}
\mat{H} = 
  \begin{bmatrix}
    1 & 0 & 0 & 0 & 0 & 0\\
    0 & 1 & 0 & 0 & 0 & 0
  \end{bmatrix}
  \equiv
  \begin{bmatrix}
    \mat{H}_1\\
    \mat{H}_2
  \end{bmatrix}
  \label{eq:Hprojection}
\end{align}
projects the target position into the 2d-plane, $\rvec{h}_i$ is uniformly distributed on a unit circle, 
while $\rvec{v}_i  \sim \gauss{0}{\covm{v}}$. $\rmat{S} \equiv \mat{S}(\rvec{p})$ describes the shape of
the target, where 
\begin{align}
  \mat{S}(\rvec{p})
  =
  \begin{bmatrix}
     l_1 \cos(\alpha) & -l_2 \sin(\alpha) \\
    l_1 \sin(\alpha) & l_2 \cos(\alpha)
  \end{bmatrix}
  \equiv
  \begin{bmatrix}
    \mat{S}_1(\vec{p})\\
    \mat{S}_2(\vec{p})
  \end{bmatrix}.
  \label{eq:definitionSi}
\end{align}

\begin{figure} 
\centering 
\def\svgwidth{0.8\columnwidth} 
\begingroup%
  \makeatletter%
  \providecommand\color[2][]{%
    \errmessage{(Inkscape) Color is used for the text in Inkscape, but the package 'color.sty' is not loaded}%
    \renewcommand\color[2][]{}%
  }%
  \providecommand\transparent[1]{%
    \errmessage{(Inkscape) Transparency is used (non-zero) for the text in Inkscape, but the package 'transparent.sty' is not loaded}%
    \renewcommand\transparent[1]{}%
  }%
  \providecommand\rotatebox[2]{#2}%
  \newcommand*\fsize{\dimexpr\f@size pt\relax}%
  \newcommand*\lineheight[1]{\fontsize{\fsize}{#1\fsize}\selectfont}%
  \ifx\svgwidth\undefined%
    \setlength{\unitlength}{227.89917143bp}%
    \ifx\svgscale\undefined%
      \relax%
    \else%
      \setlength{\unitlength}{\unitlength * \real{\svgscale}}%
    \fi%
  \else%
    \setlength{\unitlength}{\svgwidth}%
  \fi%
  \global\let\svgwidth\undefined%
  \global\let\svgscale\undefined%
  \makeatother%
  \begin{picture}(1,0.50974425)%
    \lineheight{1}%
    \setlength\tabcolsep{0pt}%
    \put(0,0){\includegraphics[width=\unitlength,page=1]{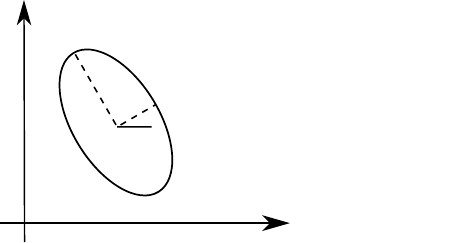}}%
    \put(0.28645444,0.24712976){\color[rgb]{0,0,0}\makebox(0,0)[lt]{\lineheight{1.25}\smash{\begin{tabular}[t]{l}$\alpha$\end{tabular}}}}%
    \put(0.34232176,0.28468299){\color[rgb]{0,0,0}\makebox(0,0)[lt]{\lineheight{1.25}\smash{\begin{tabular}[t]{l}$l_1$\end{tabular}}}}%
    \put(0.11125153,0.41220648){\color[rgb]{0,0,0}\makebox(0,0)[lt]{\lineheight{1.25}\smash{\begin{tabular}[t]{l}$l_2$\end{tabular}}}}%
    \put(0,0){\includegraphics[width=\unitlength,page=2]{measurement_model.pdf}}%
    \put(0.29470942,0.38193017){\color[rgb]{0,0,0}\makebox(0,0)[lt]{\lineheight{1.25}\smash{\begin{tabular}[t]{l}$\rvec{v}_i$\end{tabular}}}}%
    \put(0.3148946,0.46019776){\color[rgb]{0,0,0}\makebox(0,0)[lt]{\lineheight{1.25}\smash{\begin{tabular}[t]{l}$\rvec{y}_i$\end{tabular}}}}%
    \put(0.30446653,0.33359224){\color[rgb]{0,0,0}\makebox(0,0)[lt]{\lineheight{1.25}\smash{\begin{tabular}[t]{l}$\rvec{z}_i=[h_{i,1}l_1,h_{i,2}l_2]$\end{tabular}}}}%
  \end{picture}%
\endgroup%
 
\caption{This figure depicts the multiplication error measurement model for extended object tracking.  
A measurement $\rvec{y}_{i}$ is the measurement source $\rvec{z}_{i}$ corrupted with sensor noise $\rvec{v}_{i}$.  
The measurement source
$\rvec{z}_{i}$ relates object orientation $\alpha$ and axes-lengths $(l_1,l_2)^{\tp}$ by a multiplicative noise $\rvec{h}_{i}$, which are uniformly distributed on a unit
circle. }
\label{fig:measurement_model}
\end{figure}

The model is illustrated in Fig.~\ref{fig:measurement_model}. 
One of the main challenges in deriving a closed-form solution to MEM is the high non-linearity in $\rmat{S}\rvec{h}_i$.
 This hinders a simultaneous estimation of kinematics and shape using a linear minimum mean squared error (LMMSE) estimator
\cite{baumModelingTarget:2012} and motivates the introduction of a second-order pseudo-measurement 
\begin{align}
  \label{eq:ysquare}
  \rvec{Y}_i
  &= 
  \mat{F} \left( \rvec{y}_i - \evec{y}_{i} \right)\otimes\left( \rvec{y}_i - \evec{y}_{i} \right),
\end{align}
which is used to update the estimate for the shape state. Here $\vec{u} \otimes \vec{w} = \left( u_1 w_1, u_1 w_2, u_2 w_1, u_2
w_2 \right)^{\tp}$ denotes the Kronecker product. The sole purpose of the matrix $\mat{F}$ (and its sibling $\mat{\tilde F}$ which
will come in handy later) is to remove one of the identical
off-diagonals
\begin{align}
\mat{F}
&=
\left[
\begin{matrix}
  1&0&0&0\\
  0&0&0&1\\
  0&1&0&0&
\end{matrix}
\right],
\quad
\mat{\tilde F}
=
\left[
\begin{matrix}
  1&0&0&0\\
  0&0&0&1\\
  0&0&1&0&
\end{matrix}
\right].
\end{align}
It is worthwhile to note that the introduction of a second-order pseudo-measurement to account for the non-linearity can be put on
solid stochastical ground by taking a measure-theoretic viewpoint (cf.~\cite{desantisOptimalQuadratic:1995}). For a more
intuitive treatment on the basis of uncorrelated transforms we refer to \cite{lanNonlinearEstimation:2015}.

As we focus on deriving a batch measurement update, the time index $k$ is omitted for the seek of compactness.\footnote{ 
For example, $\rvec{y}_{i}$ corresponds to $\ovec{y}_{k}^{(i)}$ in
\cite{yangTrackingOrientation:2019}.} 
Given the measurement outcomes $\ovec{y}_{1}, ...,
\ovec{y}_{L}$ at our current time, the prior distributions  as $\rvec{r} \sim
\gauss{\evec{r}_0}{\covm{r}_0}$ and $\rvec{p} \sim \gauss{\evec{p}_0}{\covm{p}_0}$, our goal is to obtain a batch update along the
lines of \makebox{Eqs.~(\ref{eq:info_intro1}-\ref{eq:info_intro2})} for both kinematics and shape.

\subsection{Kinematics}
\label{sec:kinematics}
To apply the information filter to the kinematic update of the MEM-EKF*, it is necessary to identify the linearized measurement
model that has been used to derive the Kalman update given by Eqs.~(11-13) in \cite{yangTrackingOrientation:2019}. 
With
\begin{equation}
  \evec{y}_i = \mat{H} \evec{r}_{i-1},
\end{equation}
the Kalman update reads
\begin{align}
  \label{eq:kalmanBegin}
  \evec{r}_i &= \evec{r}_{i-1} + \covm{ry}_i \left[ \covm{y}_i \right]^{-1} \left( \ovec{y}_i-\evec{y}_i \right),\\
  \covm{r}_i &= \covm{r}_{i-1} - \covm{ry}_i \left[ \covm{y}_i \right]^{-1} \left( \covm{ry}_i \right)^{\tp}.
  \label{eq:kalmanEnd}
\end{align}
According to \cite{yangTrackingOrientation:2019}, $\covm{y}_i$ and $\covm{ry}_i$ are given as follows
\begin{align}
  \covm{ry}_i &= \covm{r}_{i-1} \mat{H}^{\tp},\\
  \covm{y}_i &= \mat{H}\covm{r}_{i-1} \mat{H}^{\tp} + \covm{\text{I}}_i + \covm{\text{II}}_i + \covm{v},
\end{align}
where $\covm{v}$ is the covariance of the measurement noise. Furthermore
\begin{align}
  \covm{\text{I}}_i &= \mat{S}(\evec{p}_{i-1}) \covm{h}\mat{S}(\evec{p}_{i-1})^{\tp},
\end{align}
with $\covm{h}$ being the covariance of the multiplicative noise and finally:
\begin{align}
  [\covm{\text{II}}_i]_{mn} &=  \text{tr}\left\{\covm{p}_{0}\mat{J}_{n,i-1}^{\tp}\covm{h}\mat{J}_{m,i-1}\right\}\,\text{for}\, m,n = 1,2,\\
  \mat{J}_{n,i-1}&\equiv\mat{J}_{n}(\vec{p}_{i-1})\equiv\left.\frac{\partial \mat{S}_n(\vec{p})}{\partial \vec{p}}\right|_{\vec{p}=\evec{p}_{i-1}}.
  \label{eq:definitionJi}
\end{align}
Regarding the derivation of these quantities, we refer to \cite{yangTrackingOrientation:2019}.

The Kalman update Eqs.~(\ref{eq:kalmanBegin}-\ref{eq:kalmanEnd}) can be derived from the following linearized measurement model
\begin{align}
  \rvec{y}_{i} &= \mat{H} \rvec{r} + \rvec{s}_{i},
\end{align}
where $\rvec{s}_i \sim \gauss{0}{\covm{\text{I}}_i + \covm{\text{II}}_i + \covm{v}}\equiv \gauss{0}{\covm{s}_{i}}$. In
contrast to the original model, Eq.~\eqref{eq:measModel}, the non-linearity $\rmat{S}\rvec{h}_i$ is accounted for only
effectively via a renormalization of the measurement noise. The validity of the model is easily verified by plugging the
definition of $\mathbf{H}$ from Eq.~\eqref{eq:Hprojection} and the measurement noise $\covm{s}_{i}$ into the formul\ae{} of
the Kalman update, Eqs.~(\ref{eq:kalman_linear1}-\ref{eq:kalman_linear2}).

To obtain a batch update, we approximate $\covm{s}_{i} \approx \covm{s}_{1} \equiv  \covm{s}$. This implies that we are ignoring
any feedback of changes to the object shape estimate during the kinematic update. Iterating the information update yields
\begin{align}
  \label{eq:eifkinematics1}
  \evec{\xi}^\vec{r}_{L} &= \evec{\xi}^\vec{r}_{0} + \mat{H}^{\tp} [\covm{s}]^{-1} \sum_{i=1}^L \ovec{y}_{i}, \\
  [\covm{r}_L]^{-1}     &= [\covm{r}_0]^{-1} + L \mat{H}^{\tp} [\covm{s}]^{-1} \mat{H},\\
  \evec{r}_{L}          &= \covm{r}_L \evec{\xi}^\vec{r}_{L} 
  ,\quad
  \evec{\xi}^\vec{r}_{0} = [\covm{r}_0]^{-1}\evec{r}_{0}.
  \label{eq:eifkinematics2}
\end{align}
where $L$ is the number of measurements.

\subsection{Shape Update}
\label{sec:shape}
The shape update goes along similar lines. First of all, the expectation value of the pseudo-measurement is given as
\begin{align}
  \label{eq:ysquare_expectation}
  \evec{Y}_i
  &= 
  \mat{F}\text{vect}\left\{\covm{\ovec{y}}_i\right\},
\end{align}
where the $\text{vect}\left\{\cdot\right\}$ operator transforms a matrix into a column vector as follows
\begin{align}
  \text{vect}\left\{
  \begin{bmatrix}
    m_{11} & m_{12} \\
    m_{21} & m_{22}
  \end{bmatrix}
\right\}
=
\begin{bmatrix}
  m_{11} & m_{12} & m_{21} & m_{22}
\end{bmatrix}^{\tp}.
\end{align}
The Kalman update for the shape parameters now reads (cf.~Eqs.~(25-26) in \cite{yangTrackingOrientation:2019})
\begin{align}
  \evec{p}_i &= \evec{p}_{i-1} + \covm{pY}_i \left[ \covm{Y}_i \right]^{-1} \left( \ovec{Y}_i-\evec{Y}_i \right),
  \label{eq:kalmanShape1}\\
  \covm{p}_i &= \covm{p}_{i-1} - \covm{pY}_i \left[ \covm{Y}_i \right]^{-1} \left( \covm{pY}_i \right)^{\tp},
  \label{eq:kalmanShape2}
\end{align}
with 
\begin{align}
  \covm{\vec{p}\ovec{Y}}_i = \covm{\vec{p}}_{i-1} \mat{M}_{i-1}^{\tp}
\end{align}
and (cf.~Eqs.~(\ref{eq:definitionSi}, \ref{eq:definitionJi}) for the definitions of $\mat{S}_n(\vec{p})$ and
$\mat{J}_{n}(\vec{p}_i)$)
\begin{align}
  \mat{M}_i \equiv \mat{M}(\vec{p_i}) \equiv
  \begin{bmatrix}
    2 \mat{S}_1(\vec{p}_i) \covm{\vec{h}} \mat{J}_{1}(\vec{p}_i) \\
    2 \mat{S}_2(\vec{p}_i) \covm{\vec{h}} \mat{J}_{2}(\vec{p}_i) \\
    \mat{S}_1(\vec{p}_i) \covm{\vec{h}} \mat{J}_{2}(\vec{p}_i) + \mat{S}_2(\vec{p}_i) \covm{\vec{h}} \mat{J}_{1}(\vec{p}_i)
  \end{bmatrix}^{\tp}.
  \label{eq:def_mi}
\end{align}
The covariance of the pseudo-measurement is given as
\begin{equation}
  \covm{Y}_i = \mat{F}\left(\covm{y}_i \otimes \covm{y}_i\right)(\mat{F} + \tilde{\mat{F}}).
\end{equation}
Again, we refer to  \cite{yangTrackingOrientation:2019} for the details of the derivation.

To derive a batch update based on the update formul\ae{} of the information filter, we identify the linearized measurement model 
as
\begin{align}
  \rvec{Y}_i
  &\approx
  \evec{Y}_{i}
  +
  \mat{M}_{i-1} (\rvec{p} - \evec{p}_{i-1})
  +
  \rvec{t}_i,
  \label{eq:ysquare_approx}
\end{align}
where $\rvec{t}_i \sim \gauss{0}{\covm{Y}_i - \mat{M}_{i-1}^{\tp} \covm{p}_{i-1} \mat{M}_{i-1}}\equiv
\gauss{0}{\covm{t}_i}$. Verification of correctness is again as straightforward as checking that the resulting Kalman
update for this model exactly matches Eqs.\ (\ref{eq:kalmanShape1}-\ref{eq:kalmanShape2}). Similar to the kinematic case,
additional approximations are necessary to enable a batch update. We approximate $\mat{M}_i \approx \mat{M}_0$, $\covm{t}_{i}
\approx \covm{t}_{1} \equiv \covm{t}$, $\evec{p}_{i-1} \approx \evec{p}_0$ and 
\begin{align}
  \label{eq:approxCy}
  \covm{y}_L
  &\approx 
  \mat{H} \covm{r}_{L} \mat{H}^{\tp} + \covm{\text{I}}_1+ \covm{\text{II}}_1+\covm{v},
\end{align}
such that $\evec{Y}_i = \mathbf{F}\text{vect}\left\{\covm{y}_i\right\} \approx \evec{Y}_1 \equiv \evec{Y}$. Note here that
$\covm{r}_{L}$ is already available since we perform the kinematic update first. The measurement model thus simplifies to
\begin{align}
  \rvec{Y}_i
  &\approx
  \evec{Y}
  +
  \mat{M}_{0} (\rvec{p} - \evec{p}_{0})
  +
  \rvec{t}_i,
\end{align}
with $\rvec{t}_i \sim \gauss{0}{\covm{t}}$. Also we approximate the $i$-th outcome of the pseudo-measurement as
\begin{align}
  \label{eq:approxOutcome}
  \ovec{Y}_i
  &\approx 
  \mat{F} \left( \ovec{y}_i - \evec{y}_{L} \right)
  \otimes 
  \left( \ovec{y}_i - \evec{y}_{L} \right)
\end{align}
where $\evec{y}_{L}$ is again known from the previously-run kinematic update. Iterating the information filter update yields
\begin{align}
  \evec{\xi}^\vec{p}_{L} &= \evec{\xi}^\vec{p}_{0} + \mat{M}_0^{\tp} [\covm{t}]^{-1} \sum_{i=1}^L \left[\ovec{Y}_{i} -
    \evec{Y} + \mat{M}_0 \evec{p}_0\right], 
  \label{eq:eifshape1}\\
    [\covm{p}_L]^{-1}     &= [\covm{p}_0]^{-1} + L \mat{M}_0^{\tp} [\covm{t}]^{-1} \mat{M}_0,
    \\
  \evec{p}_{L}          &= \covm{p}_L \evec{\xi}^\vec{p}_{L} 
  ,\quad
  \evec{\xi}^\vec{p}_{0} = [\covm{p}_0]^{-1}\evec{p}_{0}.
  \label{eq:eifshape2}
\end{align}
As approximating $\evec{y}_i \approx \evec{y}_L$ in Eq.~\eqref{eq:approxOutcome} and $\covm{r}_i \approx \covm{r}_L$
in Eq.~\eqref{eq:approxCy} is not the only possible choice to avoid a dependency on intermediate results from the kinematic
update, we refer to update equations Eqs.~(\ref{eq:eifshape1}-\ref{eq:eifshape2}) as MEM-EIF[$\ovec{y}_L$]. 
Another natural choice is  $\evec{y}_i \approx \evec{y}_0, \covm{r}_i \approx \covm{r}_0$ which we refer to as
MEM-EIF[$\ovec{y}_0$]. In this case, the proposed MEM-EIF[$\ovec{y}_0$] is equivalent to the MEM-EKF* if we avoid the
approximations by performing a sequential
information update, i.e., one information update per measurement. This is further discussed in the numerics part Sec.~\ref{sec:numerics}.

\subsection{Discussion of the approximations}
\label{sec:order_analysis}
\subsubsection{Overview}
As we have seen, additional approximations on top of the original update equations from \cite{yangTrackingOrientation:2019}
are required to allow for a batch update. In case of the kinematic update, these are quite intuitive: As the shape enters the
update only via a renormalization of the measurement noise $\mat{R}_{i,\vec{r}}$, approximating $\mat{R}_{i,\vec{r}} \approx
\mat{R}_{0,\vec{r}}$ should not impact the quality of the final estimate $\evec{r}_L$ too badly as long as the initial shape estimate
is reasonably close to the true value. Consequently, approximating the expected measurement $\evec{y}_i \approx \evec{y}_L =
\mat{H} \evec{r}_L$ and using $\covm{\vec{r}}_i \approx \covm{\vec{r}}_L$ for the calculation of $\covm{\ovec{y}}_i$
(cf.~Eq.~\eqref{eq:approxCy}) within the shape update seems quite reasonable. After all, $\evec{r}_L$ should be the more precise
estimate as is reflected in its lower covariance $\covm{\vec{r}}_L$ after $L$ updates.

However, regarding the remaining approximations -- $\mat{M}_i \approx \mat{M}_0$, $\covm{t}_{i} \approx \covm{t}_{1}$,
$\evec{p}_{i-1} \approx \evec{p}_0$, $\covm{\text{I}}_i \approx \covm{\text{I}}_1$ and $\covm{\text{II}}_i \approx
\covm{\text{II}}_1$ -- things are less obvious. 
Assuming the validity of the just mentioned approximation regarding the expected
measurement and its covariance, we justify the remaining approximations via an order analysis. The idea is to apply the
information update before applying any approximations, i.e., directly to the model \eqref{eq:ysquare_approx}. By construction, the
resulting information update is then
equivalent to the original Kalman update Eqs.~(\ref{eq:kalmanShape1}-\ref{eq:kalmanShape2}). We proceed to show that by
neglecting higher order corrections in the shape, the batch variant of the information update as stated in
Eqs.~(\ref{eq:eifshape1}-\ref{eq:eifshape2}) is obtained. This proves that in lowest order the batch variant of the shape update
is equivalent to its Kalman counterpart.

\subsubsection{Order analysis}

\begin{figure*}[t]
  \centering

  \begin{tikzpicture}[spy using outlines={rectangle, magnification=2,connect spies}]
    \tikzset{mark size=0.5}
    \begin{axis}
    [
      width=0.9\textwidth,
      axis equal image=true,
      legend style={fill opacity=0.8,font = \small,legend columns = 4, 
      column sep = 0.1cm, draw opacity=1, text opacity=1,at={(0.9,-0.1)}, draw=white!80!black},
      xmin=-200,xmax=5800,xlabel={$x$ [m]},
      ylabel = { $y$ [m]},ymin=-2800,ymax=2500,
      ytick={-2000,-1000,0,1000,2000},
      xtick={0,1000,2000,3000,4000,5000},
    ]
    
      \addlegendimage{area legend,black}\addlegendentry{Ground Truth}
      \addlegendimage{area legend,blue,}\addlegendentry{Random Matrix}
      \addlegendimage{area legend,color0,}\addlegendentry{MEM-EKF*}
      \addlegendimage{area legend,green!50.1960784313725!black,}\addlegendentry{MEM-EIF[$\ovec{y}_L$]}
      \input{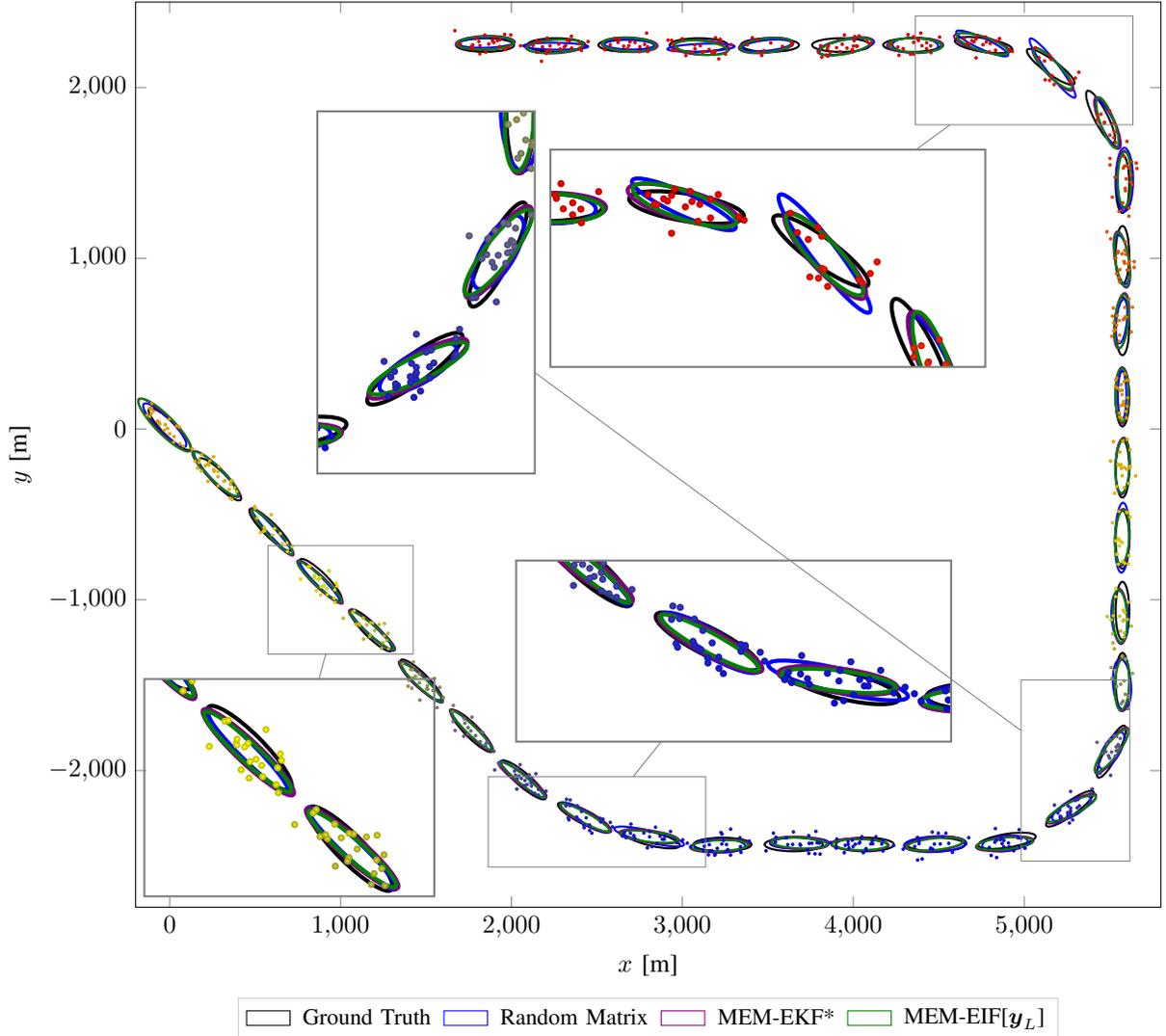} 
        
      \coordinate (spypoint1) at (axis cs:1000,-1000);
      \coordinate (magnifyglass1) at (axis cs:700,-2100);
      
      \coordinate (spypoint2) at (axis cs:2500,-2300);
      \coordinate (magnifyglass2) at (axis cs:3300,-1300);  
        
      \coordinate (spypoint3) at (axis cs:5300,-2000);
      \coordinate (magnifyglass3) at (axis cs:1500,800);
        
      \coordinate (spypoint4) at (axis cs:5000,2100);
      \coordinate (magnifyglass4) at (axis cs:3500,1000);
    \end{axis}	
       
    \spy [gray, height=3cm,  width = 4cm] on (spypoint1) in node[fill=none ] at (magnifyglass1);
    \spy [gray, height=2.5cm,width = 6cm] on (spypoint2) in node[fill=white] at (magnifyglass2);
    \spy [gray, height=5cm,  width = 3cm] on (spypoint3) in node[fill=white] at (magnifyglass3);
    \spy [gray, height=3cm,  width = 6cm] on (spypoint4) in node[fill=white] at (magnifyglass4);

	\end{tikzpicture}
	\caption{This figure shows one example run with three trackers: random matrix (blue), MEM-EKF* (red) and MEM-EIF (green). The ground truth object is black ellipses and measurements are colored dots.  }
	\label{fig:example_run}
\end{figure*}

Consider the first two information updates based on the model Eq.~\eqref{eq:ysquare_approx} where it already shows that the
$i$-dependence of $\evec{Y}_i$, $\mat{M}_{i-1}$, $\evec{p}_{i-1}$ and $\vec{t}_i$ seems to hinder the derivation of a batch
update: 
\begin{align}
  \label{eq:eifshape}
  \evec{\xi}^\vec{p}_{2} &= 
  [\covm{p}_0]^{-1} \evec{p}_{0}
    + 
    \mat{M}_{0}^{\tp}
    [\covm{t}_{1}]^{-1}
    [
    \ovec{Y}_1
    -
    \evec{Y}_1
    +
    \mat{M}_{0} \evec{p}_{0}
    ]\\
    &\quad + 
    \mat{M}_{1}^{\tp}
    [\covm{t}_{2}]^{-1}
    [
    \ovec{Y}_2
    -
    \evec{Y}_2
    +
    \mat{M}_{1} \evec{p}_{1}
    ]\nonumber.
\end{align}
If our prior is informative enough, i.e., the covariance $\covm{p}_0$ is small and $\evec{p}_0$ is close to the true state,
we can expand as follows 
\begin{align}
  \mat{M}_{1} 
  &= 
  \mat{M}(\evec{p}_1) 
  \approx
  \mat{M}(\evec{p}_0) 
  +
  \Delta \mat{M},
\end{align}
where we defined (with $(\vec{u})_k = u_k$ being the $k$-th component of the vector $\vec{u} = (u_1, u_2, u_3)^{\tp}$)
\begin{align}
  [\Delta \mat{M}]_{mn}
  \equiv
  \sum_{k=1}^3
  \left.
    \frac{\partial [\mat{M}(\vec{p})]_{ij}}{\partial p_k}
  \right|_{\vec{p} = \evec{p}_0}
  (\evec{p}_{1} - \evec{p}_0)_k.
\end{align}
Assuming $\covm{p}_0$ is small enough, we have that (with $|\cdot|$ denoting the absolute value)
\begin{align}
  |([\covm{p}_0]^{-1} \vec{p})_k| 
  &\gg 
  |(\mat{M}_0 [\covm{t}_{2}]^{-1} 
  [
    \ovec{Y}_2
    -
    \evec{Y}_2
    +
    \mat{M}_{1} \evec{p}_{1}
  ])_k|\\
  &\gg 
  |(\Delta \mat{M} [\covm{t}_{2}]^{-1} 
  [
    \ovec{Y}_2
    -
    \evec{Y}_2
    +
    \mat{M}_{1} \evec{p}_{1}
  ])_k|.
\end{align}
The first inequality above is large because the eigenvalues of $[\covm{p}_0]^{-1}$ tend to infinity in the limit $\covm{p}_0
\rightarrow 0$, while for the second we have $\evec{p}_1 - \evec{p}_0 \rightarrow 0$ in this limit. Keeping the lowest order
corrections only gives us
\begin{align}
  \evec{\xi}^\vec{p}_{2} 
  &\approx
  [\covm{p}_0]^{-1} \evec{p}_{0}
    + 
    \mat{M}_{0}^{\tp}
    [\covm{t}_{1}]^{-1}
    [
    \ovec{Y}_1
    -
    \evec{Y}_1
    +
    \mat{M}_{0} \evec{p}_{0}
    ]\\
    &\quad + 
    \mat{M}_{0}^{\tp}
    [\covm{t}_{2}]^{-1}
    [
    \ovec{Y}_2
    -
    \evec{Y}_2
    +
    \mat{M}_{1} \evec{p}_{1}
    ].\nonumber
\end{align}
Furthermore
\begin{align}
    \mat{M}_{1} \evec{p}_{1}
    \approx
    (\mat{M}_{0} + \Delta\mat{M})(\evec{p}_{0} + [\evec{p}_{1} - \evec{p}_0])
\end{align}
such that the same idea can be applied again to get
\begin{align}
  \evec{\xi}^{\vec{p}_{2}} 
  &\approx
  [\covm{p}_0]^{-1} \evec{p}_{0}
    + 
    \mat{M}_{0}^{\tp}
    [\covm{t}_{1}]^{-1}
    [
    \ovec{Y}_1
    -
    \evec{Y}_1
    +
    \mat{M}_{0} \evec{p}_{0}
    ]\\
    &\quad + 
    \mat{M}_{0}^{\tp}
    [\covm{t}_{2}]^{-1}
    [
    \ovec{Y}_2
    -
    \evec{Y}_2
    +
    \mat{M}_{0} \evec{p}_{0}
    ].\nonumber
\end{align}
On a meta level, the important insight is that terms proportional to $[\covm{p}_0]^{-1}$ are large, and therefore we can drop any
corrections proportional to $\evec{p}_1 - \evec{p}_0$ or also $\covm{p}_i$ itself as the leading order corrections are found in
zeroth-order. The other problematic quantities, $\evec{Y}_i$ and $\covm{t}_{2}$, are a bit more tedious but otherwise
straightforward as well. We have
\begin{align}
  \evec{Y}_2 &= \mat{F}\text{vect}\left\{\covm{y}_2\right\},\quad
  \covm{t}_{2}
  =
  \covm{Y}_2 - \mat{M}_{1}^{\tp} \covm{p}_1 \mat{M}_{1}
\end{align}
where
\begin{align}
  \covm{Y}_2 &= \mat{F} ( \covm{y}_2 \otimes \covm{y}_2) (\mat{F} + \mat{\tilde F}),\\
  \covm{y}_2 &\approx \mat{H} \covm{r}_{L} \mat{H}^{\tp} + \covm{\text{I}}_2+ \covm{\text{II}}_2+\covm{v}.
\end{align}
The quantity $\covm{r}_{L}$ is already known from the kinematic update which is run first. We approximate
\begin{align}
  \covm{\text{I}}_2 
  &\approx
  \mat{S}(\evec{p}_0)\covm{h}\mat{S}(\evec{p}_0)^{\tp}
  +
  \Delta \mat{S}\covm{h}\mat{S}(\evec{p}_0)^{\tp} +\mat{S}(\evec{p}_0)\covm{h}\Delta \mat{S}^{\tp}\\
  &\approx\mat{S}(\evec{p}_0)\covm{h}\mat{S}(\evec{p}_0)^{\tp}
  =\covm{\text{I}}_1,
\end{align}
with
\begin{align}
  [\Delta \mat{S}(\evec{p}_1)]_{mn}
  \equiv
  \sum_{k=1}^3
  \left.\frac{\partial [\mat{S}(\vec{p})]_{mn}}{\partial p_k}\right|_{\vec{p}=\evec{p}_0}
  (\evec{p}_{1} - \evec{p}_0)_k,
\end{align}
by keeping only zeroth-order terms as discussed above. The approximation $\covm{\text{II}}_2 \approx \covm{\text{II}}_1$ is even
more straightforward, as $\covm{\text{II}}_2$ is directly proportional to $\covm{p}_1$ and thus a negligible term. Therefore
\begin{align}
  \covm{y}_2 &\approx \mat{H} \covm{r}_{L} \mat{H}^{\tp} + \covm{\text{I}}_1+ \covm{\text{II}}_1+\covm{v} = \covm{y}_1,
\end{align}
such that $\evec{Y}_2 \approx \evec{Y}_1 \equiv \evec{Y}$ and
\begin{align}
  \covm{t}_{2}
  &=
  \covm{Y}_2 - \mat{M}_{1}^{\tp} \covm{p}_1 \mat{M}_{1}
  \approx
  \covm{Y}_1 - \mat{M}_{0}^{\tp} \covm{p}_0 \mat{M}_{0}
  =\covm{t}_{1}.
\end{align}
Putting it all together we find
\begin{align}
  \evec{\xi}^{\vec{p}_{2}} &\approx
  [\covm{p}_0]^{-1} \evec{p}_{0}
    + 
    \sum_{i=1}^{2}
    \mat{M}_{0}^{\tp}
    [\covm{t}_{1}]^{-1}
    [
    \ovec{Y}_i
    -
    \evec{Y}
    +
    \mat{M}_{0} \evec{p}_{0}
    ].
\end{align}
The same analysis applies to all further update steps $i=3,...,L$ as well which yields the final result
\makebox{Eqs.~\eqref{eq:eifshape1}-\eqref{eq:eifshape2}} by induction.

\section{Numerical Evaluation}
\label{sec:numerics}

In this section, we investigate the performance of the proposed information-filter-based batch update for the multiplicative error model (MEM-EIF).
After some remarks regarding tuning and initialization, we compare the proposed MEM-EIF with the sequential update tracker MEM-EKF*, and another batch update algorithm: \makebox{Random
Matrix (RM) \cite{feldmannTrackingExtended:2011}}. 
Next, we investigate the effect of different approximation variations in MEM-EIF as explained in Section \ref{sec:shape}.
Finally, we highlight the benefit of a batch update by comparing the runtime of MEM-EIF and its reference algorithm, MEM-EKF*.

\subsection{Tuning and Initialization }
While tuning the MEM-EIF (and to some lesser extent the MEM-EKF*) it is
important to recall that the original derivation
(cf.~\cite{yangTrackingOrientation:2019}) as well as our additional
approximations (cf.~Sec.~\ref{sec:order_analysis}) rely on a Taylor expansion around the shape
estimate. It is therefore mandatory to restrict the shape covariance to
reasonable small values. In addition, an unreasonable large covariance for the shape
parameters would put a significant amount of weight toward the unphysical negative
values of length and width. As a rule of thumb for practical applications, we
therefore recommend keeping the shape covariance below $\covm{l_1} \le
(0.4\times \bar l_1)^2, \covm{l_2} \le (0.4\times \bar l_2)^2$ based on the
current estimates for length and width $\bar l_1, \bar l_2$.

For the numerical experiments detailed below, the tracker is initialized
by calculating sample mean and covariance of the measurements at time $k=0$.
The sample mean serves as the initial position estimate while the shape estimate is
obtained as the 95\% confidence ellipse corresponding to the sample covariance.
The initial estimates for velocity and acceleration are set to zero.

\subsection{Performance along reference trajectory}
In this simulation, we consider a single extended target that moves in the 2d
plane along a reference trajectory and generates a random number of
measurements at every time step according to a Poisson distribution. The
simulated elliptical extended object has diameters of $340$ and $80$ meters and
moves with a constant velocity of $50\text{km}/\text{h}$. The measurements are
Poisson distributed with a Poisson rate $\lambda = 20$ which
is the same as in \cite{yangTrackingOrientation:2019} and
\cite{feldmannTrackingExtended:2011}. We compare our proposed MEM-EIF with the
reference MEM-EKF* and the random matrix method
\cite{kochBayesianApproach:2008, feldmannTrackingExtended:2011}. As it features the
better performance-cost ratio among the two variants, we picked the
MEM-EIF[$\ovec{y}_L$] variant over the MEM-EIF[$\ovec{y}_0$]
(see \ref{sec:evaluation_approximations} for a comparison of the two variants).
We use a constant-acceleration white-noise jerk model
\cite{rongliSurveyManeuveringTarget2003} for the prediction between the time
steps. The simulated ground truth objects, trajectory, measurements, and
estimates of three trackers from one example run are depicted  in
Fig.~\ref{fig:example_run}. Consistent with the observations in
\cite{yangTrackingOrientation:2019}, we find an increase in error of the random
random matrix approach around the three coordinated turns. The trackers based
on the multiplicative-error measurement model, on the other hand, can
constrain the size change while allowing for the necessary flexibility in the
orientation through the process noise.

\begin{figure}[t]
  \centering
  \definecolor{color0}{rgb}{0.501960784313725,0,0.501960784313725}
  \begin{tikzpicture}
  \begin{axis}
  [
    ylabel={ \footnotesize[m]},xlabel={\footnotesize $k\rightarrow$},
    width=.4\textwidth,height=1.3in,
    scale only axis,
    xmin=0,xmax=104,xmajorgrids,ymin=10,ymax=60,ymajorgrids,
    axis background/.style={fill=white},
    legend style = {at={(1,1)},font=\footnotesize, column sep = 2pt, legend columns = 3, }
	]
  \input{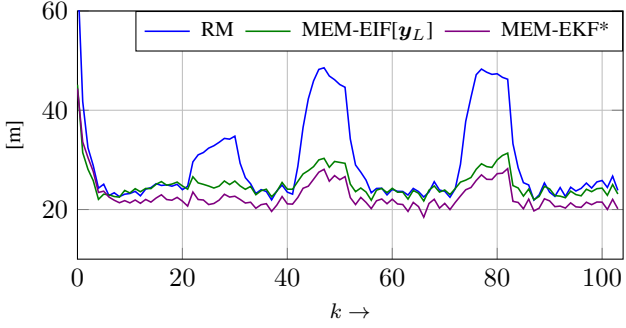}
  \end{axis}
  \end{tikzpicture}
   \caption{This figure depicts the average error for 100 simulated runs using the simulation shown in Fig.~\ref{fig:example_run}.  }
   \label{fig:error_average}
\end{figure}
We use the Gaussian Wasserstein distance \cite{yangMetricsPerformance:2016} as the metric for evaluating the estimation error of elliptical shapes. The Gaussian Wasserstein distance incorporates the estimation error on both shape and location, can be calculated in closed-form, and is a true metric. 
As the center $\mat{H}\rvec{r}$ and shape matrix $\mat{\Sigma}=\mat{S}\mat{S}^{\tp}$ of an ellipse can be interpreted as the mean and covariance matrix of a Gaussian distribution,  the Wasserstein distance between two elliptical objects is 
\begin{align}\notag
d_{GW}^2(\rvec{x}_1,\rvec{x}_2)&=||\mat{H}\rvec{r}_1-\mat{H}\rvec{r}_2||^{2}\\
&+\Trace{\mat{\Sigma}_1+\mat{\Sigma}_2-2\sqrt{\sqrt{\mat{\Sigma}_1}\mat{\Sigma}_2\sqrt{\mat{\Sigma}_1}}}.
\label{eq::distGaussianWasserstin}
\end{align}
As shown in Fig.~\ref{fig:error_average}, among the batch updates the
\makebox{MEM-EIF[$\rvec{y}_L$]} outperforms the random matrix approach at the
coordinated turns while showing a similar error otherwise. Compared to its
reference MEM-EKF*, we see an increase in error of roughly ten percent due to
the additional approximations.

\begin{figure}[t]
  \centering
  \begin{tikzpicture}
  \begin{axis}
  [
    ylabel={ \footnotesize[m]},xlabel={\footnotesize $k\rightarrow$},
    width=.4\textwidth,height=1.3in,
    scale only axis,
    xmin=0,xmax=104,xmajorgrids,ymin=10,ymax=40,ymajorgrids,
    axis background/.style={fill=white},
    legend style = {at={(1,1)},font=\footnotesize, column sep = 2pt, legend columns = 1, }
	]
  \input{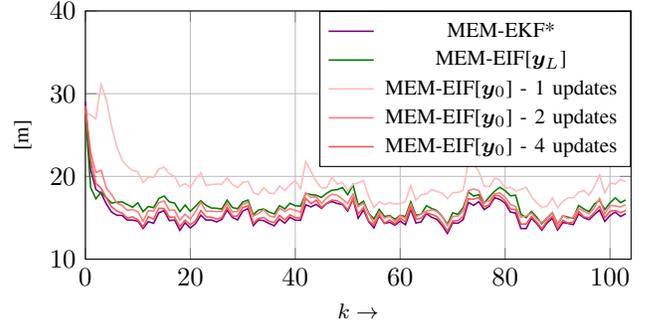}
  \end{axis}
  \end{tikzpicture}
  \caption{Estimation Error of the MEM-EIF[$\evec{y}_0$] for different batch sizes ($U$~updates is equivalent to a batch size of
  $L/U$) versus MEM-EKF* and MEM-EIF[$\rvec{y}_L$]. We used a Poisson rate $\lambda = 50$ for the simulation, otherwise the
simulation is the same as in Fig.~\ref{fig:example_run}. The results are averaged over 100 runs. }
   \label{fig:info_convergence}
\end{figure}

\subsection{Runtime performance}
\begin{figure}[t]
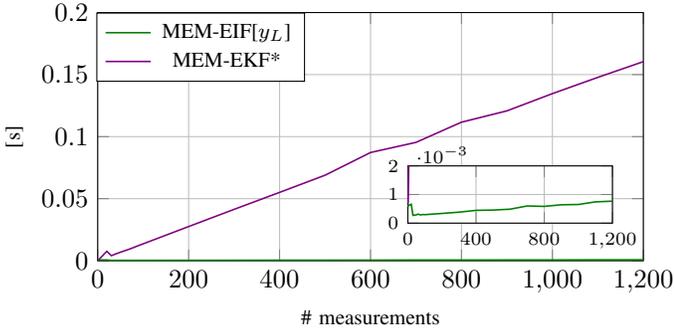

  \centering
  \begin{tikzpicture}
    \path[use as bounding box] (2,-1) rectangle (3.0, 3.5);
  \begin{axis}
  [
    yticklabel style={
      /pgf/number format/fixed,
      /pgf/number format/precision=2
    },
    scaled y ticks=true,
    ylabel={ \footnotesize[s]},
    xlabel={ \footnotesize \# measurements},
    width=.4\textwidth,height=1.3in,
    scale only axis,
    xmin=0,xmax=1200,xmajorgrids,ymin=0,ymax=0.2,ymajorgrids,
    axis background/.style={fill=white},
    legend style = {at={(0.38,1)},font=\footnotesize, column sep = 2pt, legend columns = 1, }
	]
  \input{figs/ShishanTrajectoryRuntimeComparison.tex}
  \end{axis}
  \end{tikzpicture}
  \begin{tikzpicture}
    \path[use as bounding box] (-1,-1.5) rectangle (0.4, 3.0);
  \begin{axis}
  [
    yticklabel style={
      /pgf/number format/fixed,
      /pgf/number format/precision=2
    },
    tick label style={font=\scriptsize},
    xtick distance=400,
    ytick distance=0.001,
    width=.15\textwidth,height=0.3in,
    scale only axis,
    xmin=0,xmax=1200,xmajorgrids,ymin=0,ymax=0.002,ymajorgrids,
    axis background/.style={fill=white},
	]
  \input{figs/ShishanTrajectoryRuntimeComparisonMemEIF.tex}
  \end{axis}
  \end{tikzpicture}
  \caption{Comparison of the runtime performance of the sequential MEM-EKF* (purple line) and the batch MEM-EIF[$\rvec{y}_L$] (green
    line) update based on our python implementation. The number of seconds necessary to perform the update (y-axis) is plotted
    against the number of measurements which are part of the update (x-axis). Due to the large difference in the scaling constant,
    the green line representing the scaling of the MEM-EIF[$\rvec{y}_L$] update is hugging the x-axis. In the inset, the same data is
  shown such that the linear scaling of the MEM-EIF[$\rvec{y}_L$] update in the number of measurements becomes visible as well.}
  \label{fig:runtime_performance}
\end{figure}

The main motivation for employing the batch update, i.e., MEM-EIF[$\rvec{y}_L$], instead of the sequential update is an improved runtime performance. 
Therefore,  we compare the scaling of our Python implementation of the batch update with  the sequential MEM-EKF* updates. 
The result is shown in  Fig.~\ref{fig:runtime_performance}.
As we can see,  both update schemes scale linearly in the number of measurements $L$, but the scaling constant of the batch update is roughly a hundred times smaller than that of the batch update. 
The big performance win for the batch update stems from eliminating the explicit main loop of the sequential update which performs
one Kalman update of kinematics and shape per measurement
(\makebox{Eqs.~(\ref{eq:kalmanBegin}-\ref{eq:kalmanEnd},\ref{eq:kalmanShape1}-\ref{eq:kalmanShape2})} above; see TABLE~I in \cite{yangTrackingOrientation:2019} for pseudo code). 
This loop in the sequential update cannot be easily vectorized due to the dependence of the $n$-th update on the $(n-1)$-th update and causes a huge performance penalty, while the batch update can profit from an implicit loop via vectorized summation over all measurements and pseudo-measurements. 
In principle, even a parallelization of this summation (e.g., on a GPU) would be possible here. 
Despite there being a difference in scaling constant only, we therefore expect a similarly significant speedup in general (i.e., independent of our particular Python implementation) if one chooses to switch from a sequential to a batch update. 
Finally, while we have not measured explicit performance metrics for our implementation of the random matrix approach, it should sit close the MEM-EIF[$\rvec{y}_L$] as it is also a batch update, i.e., it works based on an implicit vectorized loop over the measurements.

\subsection{Convergence of MEM-EIF[$\rvec{y}_0$] against MEM-EKF*}
\label{sec:evaluation_approximations}

As explained in \ref{sec:shape}, approximating $\evec{y}_i \approx \evec{y}_L, \covm{r}_i \approx \covm{r}_L$ in
Eq.~\eqref{eq:approxCy} is not the only natural choice. An interesting alternative is given by the approximation $\evec{y}_i
\approx \evec{y}_0, \covm{r}_i \approx \covm{r}_0$ (referred to as MEM-EIF[$\rvec{y}_0$]) since it reduces to the MEM-EKF* if we
perform one information update per measurement. This interesting property suggests the batch size, which we recall to be $L$ (the
number of measurements) for the MEM-EIF[$\rvec{y}_L$], as a tuning parameter of the MEM-EIF[$\rvec{y}_0$] approximation. It allows
for trading accuracy versus performance, with the maximal accuracy given for batch size equal to one while the maximal performance
is obtained for batch size equal to~$L$. In Fig.~\ref{fig:info_convergence} we compare the estimation error of the
\makebox{MEM-EIF[$\rvec{y}_0$]} with different batch sizes against the MEM-EKF* and the MEM-EIF[$\rvec{y}_L$]. As is apparent from
the plot, the quality of the MEM-EIF[$\rvec{y}_0$] is significantly inferior if only a single update step is performed, i.e., if
the batch size is taken to be $L$. However, by increasing the batch size to $L/2$ (implying two update steps) and $L/4$ (implying
four update steps) we see a quick improvement in the quality and, as to be expected, convergence against the MEM-EKF*. The
drawback is of course an increase in runtime and therefore it needs to be decided per application if the increase in quality is
worth the cost. In general, we recommend the MEM-EIF[$\rvec{y}_L$] over the MEM-EIF[$\rvec{y}_0$]. In this regard, it is worthwhile to note that
varying the batch size of the MEM-EIF[$\rvec{y}_L$] did not show any quality improvements in our experiments. This is due to the
fact that there is no limit in which the \makebox{MEM-EIF[$\rvec{y}_L$]} becomes equivalent to the MEM-EKF*. Therefore, the
\makebox{MEM-EIF[$\rvec{y}_L$]} should always be run with batch size equal to $L$.

\section{Remarks \& Conclusion}\label{sec:conclusion}
The need for an elliptical-target tracker typically arises in the context of a larger system which also features a computationally
demanding association step. 
One common way to reduce complexity is to work with the marginal association probabilities for each
track which in the context of point-target tracking is often combined with the probability density approximation (PDA)
\cite{bar-shalomMultitargetMultisensorTracking:1990}. 
Two extensions of this idea to extended target tracking that were recently proposed in \cite{yangLinearTimeJoint:2018,
yangMarginalAssociation:2020} involved a sequential PDA update based on the MEM-EKF*.
Our proposed batch update straightforwardly generalizes to this setting as the summation over measurements (Eq.~\eqref{eq:eifkinematics1})
and pseudo-measurements (Eq.~\eqref{eq:eifshape1}) turn into weighted summations -- the weights being the association
probabilities of each measurement. This way, the sequential PDA update can be replaced by a weighted batch update -- thus offering
the performance benefit to this important setting.

Another interesting remark concerns the observation made in  \cite{yangTrackingOrientation:2019} that the sequential update
of the MEM-EKF* depends on the order of the measurements. For the proposed batch update this is clearly not the case: the sums in
\makebox{Eqs.\ (\ref{eq:eifkinematics1}-\ref{eq:eifkinematics2},\ref{eq:eifshape1}-\ref{eq:eifshape2})} are invariant under changes
to the measurement order. Based on our analysis in Sec.~\ref{sec:order_analysis} we conclude that the order dependence stems from
the higher-order corrections in the shape update, as well as the feedback of the shape update to the kinematics. Giving up on both
yields the order-invariant batch update -- the price being a slightly increased estimation error and the benefit a significantly
reduced runtime.

To conclude this work, we like to highlight again that the newly-introduced tracker MEM-EIF[$\vec{y}_L$] combines two important strengths of
the random matrix and the MEM-EKF* approach: Similar to the random matrix approach it is a batch update, i.e., the target state is
updated in a single step using only the sums over all measurements and the MEM-specific pseudo-measurements. Compared to a
sequential measurement-by-measurement update, the scaling constant is drastically reduced leading to a strong increase in
performance. At the same time, as
the tracker is based on the MEM-EKF*, it offers direct access to the tracked ellipse parameters and their covariance thus easing
the application of relevant motion models.

\section*{Acknowledgments}
The authors would like to thank Marcus Baum for helpful suggestions on an early draft of this manuscript. Christian Gramsch thanks
Clemens Wuth for pair-debugging sessions and Stephan Timmer for proof reading the manuscript.

\begin{appendices}
\section{Derivation of the Information Filter}
\label{ap:info_derivation}
For completeness, we give a possible derivation of the information filter in this appendix. We assume a linear measurement model
as stated in Eq.\ \eqref{eq:linearmodel}. Based on the Kalman update in Eqs.~(\ref{eq:kalman_linear1}-\ref{eq:kalman_linear2}) we
define $\evec{\xi}_i = [\covm{x}_i]^{-1} \evec{x}_i$ such that
\begin{align}
  \evec{x}_{i} &= \evec{x}_{i-1} + \covm{xy}_i[\covm{y}_i]^{-1}(\ovec{y}_i - \mat{H} \evec{x}_{i-1})\\
 &= \covm{x}_{i-1} \evec{\xi}_{i-1} + \covm{xy}_i[\covm{y}_i]^{-1}(\ovec{y}_i 
                   - \mat{H} \covm{x}_{i-1} \evec{\xi}_{i-1})\\
  &= \left[
                      \covm{x}_{i-1} - \covm{x}_{i-1} \mat{H}^{\tp} [\covm{y}_{i-1}]^{-1} \mat{H} \covm{x}_{i-1} 
                   \right] 
                   \evec{\xi}_{i-1}\\
                &\phantom{=}\ + \covm{x}_{i-1} \mat{H}^{\tp} [\covm{y}_{i-1}]^{-1} \ovec{y}_i\nonumber\\
                &= \covm{x}_{i} \evec{\xi}_{i-1} + \covm{x}_{i-1} \mat{H}^{\tp} [\covm{y}_{i-1}]^{-1} \ovec{y}_i
\end{align}
which gives the intermediate result
\begin{align}
  \evec{\xi}_{i} = \evec{\xi}_{i-1} + [\covm{x}_{i}]^{-1}\covm{x}_{i-1}\mat{H}^{\tp} [\covm{y}_{i-1}]^{-1} \ovec{y}_i.
\end{align}
From the matrix inversion lemma we have the relation
\begin{align}
  [\covm{x}_{i}]^{-1}
  &=
   \left[
     \covm{x}_{i-1} - \covm{x}_{i-1} \mat{H}^{\tp} [\covm{y}_{i-1}]^{-1} \mat{H} \covm{x}_{i-1} 
   \right]^{-1}\\
  &=
  [\covm{x}_{i-1}]^{-1} + \mat{H}^{\tp} [\covm{v}]^{-1} \mat{H},
\end{align}
which gives us
\begin{align}
  [\covm{x}_{i}]^{-1} &\covm{x}_{i-1} \mat{H}^{\mathbf T} [\covm{y}_{i-1}]^{-1}\nonumber\\
  &= 
  \left[ 
    \covm{x}_{i-1} + \mat{H}^{\tp} [\covm{v}]^{-1} \mat{H}
  \right] 
  \covm{x}_{i-1} \mat{H}^{\tp} [\covm{y}_{i-1}]^{-1}\\
  &=
  \left[ 
    \mat{H}^{\tp} + \mat{H}^{\tp} [\covm{y}]^{-1} \mat{H} \covm{x}_{i-1} \mat{H}^{\tp}
  \right] 
  [\covm{y}_{i-1}]^{-1}\\
  &=
  \mat{H}^{\tp} [\covm{v}]^{-1}
  \underbrace{
    \left[ 
      \covm{v} + \mat{H} \covm{x}_{i-1} \mat{H}^{\tp}
    \right] 
  }_{=\covm{y}_{i-1}} [\covm{y}_{i-1}]^{-1}\\
  &= \mat{H}^{\tp} [\covm{v}]^{-1}
\end{align}
The information filter update is thus found to be
\begin{align}
  \evec{\xi}_{i} &= \evec{\xi}_{i-1} + \mat{H}^{\tp} [\covm{v}]^{-1} \ovec{y}_i,\\
  [\covm{x}_{i}]^{-1} &= [\covm{x}_{i-1}]^{-1} + \mat{H}^{\tp} [\covm{v}]^{-1} \mat{H}.
\end{align}

\begin{table}
  \caption{Measurement update of the MEM-EIF[$\rvec{y}_L$] algorithm}
\label{tab:pseudo_code_meas}
\rule[0pt]{\columnwidth}{1pt}
\textbf{Input:}  Measurement outcomes $\{\ovec{y}_{1}, \dots, \ovec{y}_L\}$, predicted estimates  $\evec{r}_0$,  $\evec{p}_{0}$,
$\covm{r}_0$, $\covm{p}_0$, measurement noise covariance $\mat{C}^{\rvec{v}}$, $\mat{H}$ as defined in \eqref{eq:measModel},
multiplicative noise covariance $\covm{h}$\\
\textbf{Output:}  updated estimates $\evec{r}_L ,\;  \evec{p}_L  \text{ and }  \covm{r}_L, \;  \covm{p}_L$
\begin{align*}
\text{Helper quantities: }\\
\left[
  \begin{matrix}
    \alpha & l_1 & l_2
  \end{matrix}
\right]^{\tp}
&=\evec{p}_0\\ \mat{S} =
\left[
\begin{matrix}
  \mathbf{S_1}\\\mathbf{S_2}
\end{matrix}
\right]
&=
\left[
\begin{matrix}
  \cos{\alpha} & -\sin{\alpha}\\
  \sin{\alpha} & \cos{\alpha}
\end{matrix}
\right]
\left[
\begin{matrix}
  l_1&0\\
  0&l_2
\end{matrix}
\right]\\
\mathbf{J_1}&=
\left[
\begin{matrix}
  -l_1\sin{\alpha} &\cos{\alpha}&0\\
  -l_2\cos{\alpha}&0&-\sin{\alpha}
\end{matrix}
\right]
\\
\mathbf{J_2}&=
\left[
\begin{matrix}
  l_1\cos{\alpha}&\sin{\alpha}&0\\
  -l_2\sin{\alpha}&0&\cos{\alpha}
\end{matrix}
\right]\\
  \covm{\text{I}} &= \mat{S} \covm{h}\mat{S}^{\tp}\\
  \covm{\text{II}} = [\epsilon_{mn}]  &=  \text{tr}\left\{\covm{p}_{0}\mat{J}_n^{\tp}\covm{h}\mat{J}_m\right\}
\text{ for }m,n=1,2\\
\mathbf{M}&=
\left[ 
\begin{matrix}
	2\mathbf{S_1}\mathbf{C}^{{h}}\mathbf{J_1}\\
	2\mathbf{S_2}\mathbf{C}^{{h}}\mathbf{J_2}\\
	\mathbf{S_1}\mathbf{C}^{{h}}\mathbf{J_2}+\mathbf{S_2}\mathbf{C}^{{h}}\mathbf{J_1}
\end{matrix}
\right]\\
\mathbf{F}
&=
\left[
\begin{matrix}
  1&0&0&0\\
  0&0&0&1\\
  0&1&0&0&
\end{matrix}
\right],\quad
\widetilde{\mathbf{F}}
=
\left[
\begin{matrix}
1&0&0&0\\
0&0&0&1\\
0&0&1&0&
\end{matrix}
\right]\\
\text{Kinematic Update: }\\
\covm{s} &= \mathbf{C}^{\text{I}} + \mathbf{C}^{\text{II}} + \mathbf{C}^{v}\\
\covm{r}_{L} &= \left[\covm{r}_{0} + L \mat{H}^{\tp}
[\covm{s}]^{-1}\mat{H}\right]^{-1}\\
\evec{\xi}^\vec{r}_0 &= [\covm{r}_0]^{-1} \evec{r}_0\\
\evec{\xi}^\vec{r}_L &= \evec{\xi}^\vec{r}_0 + \mat{H}^{\tp} [\covm{s}]^{-1} \sum\nolimits_{i=1}^L \ovec{y}_i\\
\evec{r}_L &= \covm{r}_L \evec{\xi}^\vec{r}_L\\
\text{Shape Update: }\\
\covm{y} &= \mat{H} \covm{r}_{L} \mat{H}^{\tp} + \covm{\text{I}} + \covm{\text{II}} + \covm{v}\\
\covm{Y} &= \mat{F}(\covm{y}\otimes\covm{y})(\mat{F}+\widetilde{\mat{F}})^{\tp}\\
\covm{t} &= \covm{Y} - \mat{M}^{\tp} \covm{p}_{0} \mat{M}\\
\covm{p}_{L} &= \left[\covm{p}_{0} + L \mat{M}^{\tp}
[\covm{t}]^{-1}\mat{M}\right]^{-1}\\
  \ovec{Y}_i &= \mat{F}\left((\ovec{y}_{i}-\mat{H} \evec{r}_L)\otimes(\ovec{y}_{i}-\mat{H} \evec{r}_L)\right)\\
\evec{Y} &= \mat{F}\text{vect}\left\{\covm{y}\right\}\\
\evec{\xi}^\vec{p}_0 &= [\covm{p}_0]^{-1} \evec{p}_0\\
\evec{\xi}^\vec{p}_L &= \evec{\xi}^\evec{p}_0 + \mat{M}^{\tp} [\covm{t}]^{-1} 
\left(
  \sum\nolimits_{i=1}^L
  \ovec{Y}_i
  - 
  L \evec{Y}
  +
  L \mat{M}\vec{p}_0 
\right)\\
\evec{p}_L &= \covm{p}_L \evec{\xi}^\vec{p}_L
\end{align*} 
\rule[0pt]{\columnwidth}{1pt}
\end{table}
\end{appendices}

\bibliographystyle{IEEEtran}
\bibliography{zotero}

\end{document}